\documentclass{pasa}%
\usepackage{color}
  
\def\etal{{\rm et~al.~}} 
\def\lapp{\ifmmode\stackrel{<}{_{\sim}}\else$\stackrel{<}{_{\sim}}$\fi}
\def\gapp{\ifmmode\stackrel{>}{_{\sim}}\else$\stackrel{>}{_{\sim}}$\fi}        

\title[DECamERON]{Superluminous supernovae at high redshift}
\author[Abbott et al.]{Tim Abbott$^1$, Jeff Cooke$^{2,3}$, Chris Curtin$^{2,3}$, Shahab Joudaki$^{2,3}$, Antonios Katsianis$^{4
}$, Anton Koekemoer$^5$, Jeremy Mould$^{2,3}$, Edoardo Tescari$^{3,6}$, Syed Uddin$^{2,7}$, Lifan Wang$^8$
\\
\\
\affil{$^1$Cerro Tololo Interamerican Observatory}
\affil{$^2$Centre for Astrophysics \& Supercomputing, Swinburne University}
\affil{$^3$ARC Centre of Excellence for All-sky Astrophysics (CAASTRO)}
\affil{$^4$Dept of Astronomy, Universidad de Chile, Camino El Observatorio 1515, Las Condes, Santiago, Chile}
\affil{$^5$Space Telescope Science Institute}
\affil{$^6$School of Physics, University of Melbourne}
\affil{$^7$Purple Mountain Observatory, Chinese Academy of Sciences, Nanjing, China}
\affil{$^8$Texas A\&M University}}

\begin{document}

\begin{abstract}
Superluminous supernovae are beginning to be discovered at redshifts as early as the epoch
of reionization. A number of candidate mechanisms is reviewed, together with the discovery programs.
\end{abstract}


\maketitle

\section{Introduction}

An unavoidable problem in probing the high redshift Universe is the rapid rise of luminosity distance
 with redshift. In luminosity distance, the epoch of reionization (EoR) from t = 100 Myrs extends 
80 Hubble radii. 
More powerful telescopes and instruments are needed for the EoR, such as:  
(i) JWST: [powerful, but small field, with spectroscopy paramount]
(ii) DECam: [1$\mu$m and shorter, wide field (Flaugher \etal 2012)]
(iii) KDUST: [1$\mu$m $<~\lambda~<$ 3$\mu$m, wide field (Yuan \etal 2012)]
(iv) Subaru: [Hyper Suprime Cam (HSC -- Miyazaki 2015)].
In this paper, we review 
survey successes and expectations for superluminous supernovae (SLSNe). 

\section{Massive and supermassive stars}
We observe very massive stars ($>$100 M$_\odot$) in the present universe and, as a result, we do know that they can form.  
But how do they actually end their lives?  Will they die as core-collapse supernovae (SNe) energized by the spin 
down of magnetars, pair instability supernovae (PISNe) with explosion energies up to 100 times that of regular 
supernovae, pulsational pair instability supernovae that can produce very bright events due to colliding shells, or via another mechanism?  Asymmetry may play a critical role in evolution. Extreme events that trigger relativistic jets may become luminous enough such that their discovery can be made by relatively small telescopes. 
 These extreme cases may output high energy radiation as in gamma-ray bursts (GRB).

Woosley \& Heger (2015) reviewed the theory of the evolution and death of  stars  heavier than 10 M$_\odot$
on the main sequence. The more massive of these, absent serious mass loss, either make black holes when they die, 
or, for helium cores exceeding $\sim$35 M$_\odot$, encounter the pair instability. Outcomes, 
including the appearance of GRBs (Levan et al 2016), depend on the initial composition of the  star, its rotation rate, 
and detailed physics. These stars can produce some of the  brightest  SNe, but also some of the faintest.

Yoshida et al (2014) investigated very  massive  stars  with main sequence mass larger than 100 M$_\odot$
and  metallicity 0.001 $<$ Z $<$ 0.004 which might explode as Type Ic  SLSNe.  
Progenitors of 43 and 61 M$_\odot$ WO  stars  with Z = 0.004 were evolved from initial 110 and 250 M$_\odot$
stars. These stars were  expected to explode as Type le  SNe. Other progenitor spectral types were studied 
by Groh et al (2013). Dessart et al (2012) point out that mixing challenges the ability to infer progenitor and explosion
properties.

From the collapse of supermassive stars, supermassive black holes observed at high redshift in QSOs could grow from direct collapse black holes 
with mass $\sim$10$^5$M$_\odot$.
Ultra-luminous supernovae (Matsumoto \etal 2016) of $\sim$10$^{45-46}$ erg s$^{-1}$ 
 would be detectable by future telescopes in the near infrared, such as, Euclid, WFIRST, KDUST and JWST for $\sim$5000 
 days to z $\lapp$ 20 and $\sim$100 events per year.

 The unknowns in binary massive star evolution
have recently received widespread attention with the detection of a
massive binary inspiral (Abbott \etal 2016a).
Understanding such events in the low-redshift Universe will enable us to better interpret high redshift observations.

\section{High z SNe} 
Does the collapse of pristine gas in the early Universe lead to the formation of very massive stars ? 
Larson (1998), Heger \&  Woosley (2002), O'Shea \& Norman (2007, 2008) 
have considered the possibility and the notion of a different initial mass function (IMF) from today's. Wide-area, deep
surveys are seeing a rare class of SLSNe, 10 -- 100 times more luminous than typical SNe 
(Quimby et al. 2011; Gal-Yam 2012).
 However, only $\gapp$50 SLSNe have been detected at low z 
(Nicholl et al 2015, Smith et al. 2007; GYL 2009; Pastorello et al. 2010; Gezari et al. 2009). One of these 
events with slow fade was thought to be powered by the radiative decay of $^{56}$Ni (Gal-Yam et al  2009). 
A single event has been identified as the first detection of 
a third type of SN with a pair-instability supernova (PISN).  As discussed above, 
SLSNe might to occur with higher rate at earlier times, due to the presence of 
pristine gas and a top-heavy IMF, which favours the creation of massive stars. 
Overall, according to Tescari et al  (2014) and 
Katsianis et al (2015) an efficient feedback mechanism is needed to obtain 
the observed star formation rate functions and stellar mass functions at 
high redshifts and SLSN maybe could play a major role.
 
Physical models of SLSNe include pair instability supernovae (PISNe), magnetars, quark novae, radiatively shocked circumstellar matter, and jet-cocoon structures. 
The energy output may be as high as 10$^{55-56}$ ergs, exceeding the main sequence radiated energy of 100 M$_\odot$ stars at 10$^{54}$ ergs.
Models involving more than one of the concepts outlined below have been considered by Tolstov et al (2016), Gal-Yam (2016) and Gilmer et al (2016).

\subsection{The PISN concept}

PISNe have been theorized since the 1960s (Rakavy \& Shaviv 1967; Barkat et al 1967) as the
 result of the deaths of stars with progenitor masses of 140--260 M$_\odot$  (Heger \& Woosley 2002; Kasen et al 2011).
Stars this massive generate conditions in their cores that enable efficient conversion of $\gamma$-ray photons into 
electron-positron pairs, followed by
rapid conversion of pressure-supporting radiation into rest mass, violent contraction and run-away thermonuclear explosion, 
obliterating the star (Kozyreva et al 2017; Chatzopoulos et al 2015).

A number of events are candidates for PISN including SN2007bi (Gal-Yam et al 2009).  For example, Lunnan et al (2016) discuss a number of  07bi-like PISN candidates, and they have rise photometry, one of the key discriminants.  The two high redshift SLSNe of Cooke et al (2012) are perhaps the most robust PISN candidates known, in particular, the redshift 2 event.

\subsection{Magnetars, quark novae}
Energy injection by a magnetar with a rapid rotation rate and magnetic field of 0.1--1 $\times$ 10$^{14}$ G 
may supply excess luminosity.  
~Chatzopoulos et al (2016) argue that this requires fine-tuning and extreme parameters for the magnetar, 
as well as the assumption of efficient conversion of magnetar energy into radiation.

Ouyed et al (2016) show that a Quark-Nova (the explosive transition of a neutron star to a quark star) occurring a few days following the SN explosion of an oxygen Wolf-Rayet star can account for SLSNe, 
including extreme energetics and a double-peaked light-curve. The expanding remnant is used to harness the kinetic energy ($>$10$^{52}$ ergs) of the ejecta. 

\subsection{Radiatively shocked circumstellar matter and jet-cocoon structures }
Blinnikov (2016) reviews calculations, not only of the magnetar model and PISNe, but also
models explaining SLSN events with the minimum energy budget,
involving  multiple  ejections  of  mass  in  presupernova  stars.   
The  radiative shocks produced in collisions of those shells may provide the required power.
This  class  of  the  models  he  refers  to  as  ``interacting"  supernovae.

Matsumoto et al (2016) consider supermassive black holes at high redshift growing from direct collapse black holes (DCBHs) with masses $\sim$10$^5$M$_\odot$,  resulting from the collapse of supermassive stars (SMSs). 
If a relativistic jet is launched from a DCBH, then it can break out of the collapsing SMS and produce a GRB. 
Although most GRB jets may miss our line of sight, they show that the energy injected from the jet into a cocoon is huge $\sim$10$^{55-56}$ erg, so that the cocoon fireball is observed as 
an ultra-luminous supernova of $\sim$10$^{45-46}$ erg/s.

\section{Observing High redshift SLSNe}

Two SLSNe at z $\sim$ 2 have been observed: a slow evolving PISN event (Cooke et al. 2012) 
and another SLSN-I type event. 
Other surveys for high z SLSNe include the ``All-Sky Automated Survey for Supernovae" (ASAS-SN; Brown \& Warren-Son Holoien 2016),
the Palomar Transient Factory (Perley et al 2016), Subaru HSC surveys (Tanaka et al 2016) and GAIA (Staley \& Fender 2016).
The   SUperluminous  Supernova  Host galaxIES (SUSHIES) survey (Schulze et al 2016) aims to provide constraints 
on the progenitors of  SLSNe by understanding the relationship to their host galaxies. 
Appendix A  reports on the DECam Deep Fields program.   
Mould et al (2017) report the discovery of a z $\approx$ 6 SLSN in the NSF field.

\subsection{SLSN rates}
The rate of core collapse SNe is proportional to the star formation rate and affected by the IMF.  
Many studies have been made at z $<$ 0.3 (e.g., Bazin \etal 2009); few at higher z.
Most core collapse SNe are too faint, although  luminous type IIn and SLSNe are an exception.  
SLSNe are extremely luminous in the UV (Brown 2016, Yan et al 2016)
, whereas type Ia are not. 
Therefore, objects are expected to be detectable at high z.
Cooke et al (2012) suggest that, at z $\sim$ 2--4, the SLSN rate  is $\sim$4 x 10$^{-7}$ /Mpc$^3$/yr. 
This is $\lapp$0.1\% of the total core collapse SN rate at 0.9 $<$ z $<$ 1.3 
found by Dahlen et al (2012). 
The z dependence of the SLSN rate has been predicted by Tanaka \etal (2012).
At this rate, the surface density of SLSNe is given in column (2) of Table 1.
For DECam these objects are quite faint by redshift 4 (see column 3), but not beyond reach.

\vspace*{0.25 truecm}

\leftline{\bf Table 1: Expected number of SLSNe/sq deg}
\begin{tabbing}
Redshift interval \= ssssssssssssssss \=   m-M-20\kill
Redshift \> \# \>   m-M-20\\
interval \> \>   -2.5$\log$(1+z)\\
 z\>(deg$^{-2})$    \>   (mag)\\
(1)\>(2)\>(3)\\
  (1.0, 1.5)  \> 2        \>  23.8\\
  (1.5, 2.0)  \> 2.5      \>  24.5\\
  (2.0, 2.5)  \> 2.5      \>  25.0\\
  (2.5, 3.0)  \> 2.5      \>  25.4\\
  (3.0, 3.5)  \> 2.5      \>  25.7\\
  (3.5, 4.0)  \> 2.5      \>  25.9\\
  (4.0, 4.5)  \> 2.5      \>  26.1\\
\end{tabbing}

\section{Other DECam results}

\subsection{GW150914}
We observed the Prime Field (Table 2) 107 days after LIGO's first detection of a binary black hole inspiral (Abbott \etal 2016a).
The error box for GW150914 includes the Prime Field. The brightest objects present after the event and not present in 2012, 2013 with colours that exclude flare stars have $z$ $\approx$ 17.5
$i.e.$ M$_z$ $\approx$ --20.5 at the distance of the event. However, the binary black hole merger model predicts no electromagnetic counterpart for the event (Abbott et al 2016b).
A brightening of at least 7.5 mag was generally seen for these objects, which were most likely SNe.

\subsection{Large Scale Structure}
Our first two nights of the DECamERON project yielded data on the Prime field.
Candidate $i$ band dropouts have z $\gapp$ 6 and their structure across the 3 square degree field is far from uniform (Figure 1).
Details are given by Mould (2013). The colour-magnitude diagram  was shown by Mould (2015). 
Figure 1 is similar to that of Barone-Nugent \etal (2014). Reionization lights up the gas on comoving scales of a few Mpc and more
(Geil et al 2016). 
This is accessible to the DECam Deep Field project.
\begin{figure}
\begin{center}
\includegraphics[scale=.35]{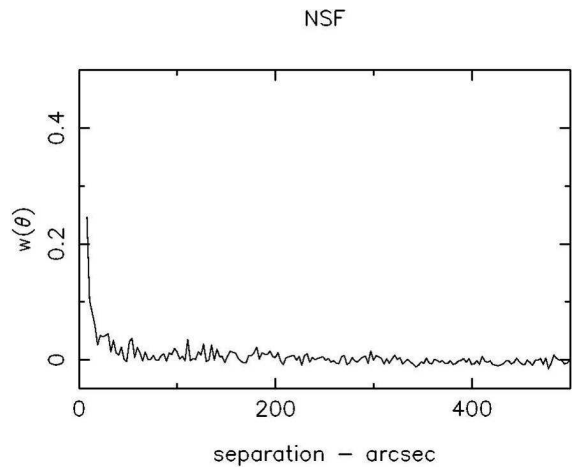}
\caption{Angular correlation function for the NSF field. The scale at redshift 6 is 5.8 kpc/arcsec.}
\end{center}
\end{figure} 

\subsection{Detection of SLSNe}
Two of us (CC \& JC) are working with 
SUDSS  (Scovacricchi \etal 2016), using Lyman Break Galaxy (LBG) colour cuts for z = 3.
For present purposes, we modified these for higher redshift.
The NSF field (Table 2) was drawn from SUDSS to increase the cadence.
For z = 3.5--4.8, these cuts are $g-r ~>~ r-i$ + 0.8;
$g-i ~>$ 0.3; --0.7 $<~ r-i ~<$ 1.2; --0.4 $<~ i-z ~<$ 0.2. These need further refinement.
Figures A1--3 (Appendix A) show our z = 4 candidates.
Mould et al (2017) report the discovery of a z $\approx$ 6 SLSN in the 
NSF field 
based upon bandpass redshifts in the table below.

\pagebreak
\leftline{\bf Table 3: Dropout bandpass vs redshift}
\begin{tabbing}
noindent $\lambda_{eff}$/(1+z) \= 100 nm\kill
\noindent $\lambda_{eff}$/(1+z)\\
----------------------------\\
 100/(1+0)   \>   100 nm\\
 360/(1+2.6) \>   $u$\\
 480/(1+3.8) \>   $g$\\
 670/(1+5.7) \>   $r$\\
 820/(1+7.2) \>   $i$\\
\end{tabbing}

\section{Higher z (towards the Dark Ages)}
There is absolutely no observational data on the stellar  Universe at z $\sim$ 20.
Events at z $>$ 20 need near infrared K-band observations. 
For this purpose, we can call on the low background and remarkable isoplanatism
in Antarctica (Aristidi \etal 2013) to conduct wide field K-band time domain surveys.
Spectroscopically, by the end of the decade we shall have JWST to detect some of these 
SLSNe at redshifts all the way to z $\sim$ 20. 
While the explosion of first generation stars is of fundamental importance, 
the rate of SLSNe at z $\sim$ 20 is highly uncertain and
a low rate  may limit the discovery space of JWST. 
The deep NIR survey to K = 29th mag proposed by Wang
\footnote{ z Equals Twenty from Antarctica (zETA) program.} (2009) would be an  excellent target feeder for JWST.

\section{Conclusions}
What we have seen in this brief review is a recent proliferation of SLSNe and a number of mechanisms that could be at work making them.
The first four years of the DECamERON project suggest that DECam, like HSC,
can penetrate the EoR. Large scale structure data are consistent with
that of other programs. There is every expectation that LBG and dropout selection criteria
will allow EoR SLSNe to be found. Time dilation of high z light curves allows
economical observing. We plan to press on to investigate the fascinating
scientific questions posed by the evolutionary end points of massive stars in the EoR. 
Among these are the possibility of using SLSNe for cosmology (Scovacricchi et al 2016).
The best route at higher z may be to search for `orphan' SLSNe that have the correct colours 
and then use very deep (JWST) spectroscopy to confirm the redshift.

\acknowledgements
We are grateful to the DES team for building DECam and the NVO at NOAO
for flattening and stacking our data. 
Parts of this research were conducted by the Australian Research Council Centre of Excellence for All-sky Astrophysics (CAASTRO), through project number CE110001020. In addition to DAOPHOT, we used
IRAF, a product of NOAO, which is operated by AURA and supported by NSF.
Thanks also to our colleagues Stuart Wyithe,
Chris Lidman,
Michele Trenti,
Robert Barone Nugent and
Andrea Kunder.

The DECamERON project used data obtained with the Dark Energy Camera, which was constructed by the Dark Energy Survey (DES) collaboration. Funding for the DES Projects has been provided by the DOE and NSF (USA), MISE (Spain), STFC (UK), HEFCE (UK). NCSA (UIUC), KICP (U. Chicago), CCAPP (Ohio State), MIFPA (Texas A\&M), CNPQ, FAPERJ, FINEP (Brazil), MINECO (Spain), DFG (Germany) and the collaborating institutions in the Dark Energy Survey, which are Argonne Lab, UC Santa Cruz, University of Cambridge, CIEMAT-Madrid, University of Chicago, University College London, DES-Brazil Consortium, University of Edinburgh, ETH Zurich, Fermilab, University of Illinois, ICE (IEEC-CSIC), IFAE Barcelona, Lawrence Berkeley Lab, LMU Munchen and the associated Excellence Cluster Universe, University of Michigan, NOAO, University of Nottingham, Ohio State University, University of Pennsylvania, University of Portsmouth, SLAC National Lab, Stanford University, University of Sussex, and Texas A\&M University. 

\section*{References}
Abbott, B. \etal 2016a, Phys Rev Lett, 116, 1102\\
Abbott, B. \etal 2016b, ApJ, 826, L13\\
Aristidi, E. \etal 2013, IAU Symposium 288, 300\\
Barone-Nugent, R. \etal 2014, ApJ, 793, 17\\
Barkat, Z., Rakavy, G. \& Sack, N 1967, PRL, 18, 379\\
Bazin, G. \etal 2009, A\&A, 499, 653 \\
Blinnikov, S. 2016, arxiv, 1611.00513\\
Brown, J. \& Warren-Son Holoien, T.	 2016, AAS, 227, 34913\\
Brown, P. 	 2016, AAS, 227, 23707\\	
Chatzopoulos, E. et al 2015, ApJ, 799, 18\\
Chatzopoulos, E. et al 2016, ApJ, 828, 94\\
Cooke, J., \etal 2012, Nature, 491, 228\\
Dahlen, T. et al 2012, ApJ, 757, 70\\
Dessart, L. \etal 2012, MNRAS, 424, 2139\\
Flaugher, B. 2012, SPIE, 8446, 11\\
Gal-Yam, A. 2012, Science, 337, 927\\
Gal-Yam, A. \& Leonard, D. 2009, Nature, 458, 865\\
Gal-Yam, A. \etal 2009, Nature, 462, 624\\
Gal-Yam, A. 2016, arxiv 1611.09353\\
Geil, P. \etal 2016, MNRAS, 462, 804\\	
Gezari, S., \etal 2009, ApJ, 690, 1313\\
Gilmer, M. et al 2016, arxiv 1610.00016\\
Groh, J. \etal 2013, A\&A, 558, 131\\
Heger, A. \& Woosley, S. E. 2002, ApJ, 567, 532\\%
Kasen, D., Woosley, S., \& Heger, A., 2011, ApJ, 734, 102\\
Katsianis, A. \etal 2015, MNRAS, 448, 3001\\
Kozyreva, A. \etal 2017, MNRAS, 464, 2854 \\
Larson, R. B., 1998, MNRAS, 301, 569 \\
Levan, A. \etal 2016, SSRv, 202, 33\\
Lunnan, R. \etal 2016, ApJ, 831, 144\\	
Matsumoto, T. et al 2016, ApJ, 823, 83\\
Miyazaki, S. 2015, IAUGA, 225, 5916\\
Moriya, T., et al. 2016, ApJ, 833, 64\\
Mould, J. 2015, ASP Conf. Ser, 491, 215\\
Mould, J. et al 2017, submitted to Chinese Science Bulletin\\
Mould, J. 2013, http://www.caastro.org/reionization-in-the-red-centre-presentations\\
Nicholl, M. et al 2015, MNRAS, 452, 3869\\
O'Shea, B. \& Norman, M., 2007, ApJ, 654, 66 \\
O'Shea, B. \& Norman, M., 2008, ApJ, 648, 31\\
Ouyed, R. et al 2016, ApJ, in press, arxiv 1611.03657\\
Pastorello, A., \etal 2010, ApJ, 724, 16\\
Perley, D. et al	 2016, ApJ, 830, 13\\	
Quimby, R. \etal 2011, Nature, 474, 487\\
Rakavy, G. \& Shaviv, G., 1967, ApJ, 148, 803\\
Scovacricchi, D. et al 2016, MNRAS, 456, 1700\\
Smith, N., \etal 2007, ApJ, 666, 1116\\
Staley, T. \& Fender, R.	 2016, arXiv 1606.03735\\	
Tanaka, M. \etal 2012, MNRAS, 422, 2675\\
Tanaka, M. \etal 2016, ApJ, 819, 5	\\
Tescari, E. \etal 2014, MNRAS, 438, 3490\\
Tolstov, A. \etal 2016, arxiv 1612.01634\\
Trenti, M \etal~2010, ApJ, 714, 202 \\ 
Valdes, F. \etal 2014, ASP Conference Series 485, 379\\
Wang, Lifan 2009, AAS, 213, 22604\\
Woosley, S. \& Heger, A. 2015, ASSL, 420, 199\\
Yan, Lin \etal 2016, arXiv, 1611.02782\\
Yoshida, T. \etal 2014, AIPC, 1594, 284\\
Yuan, Xiangyan \etal 2013, IAUS, 288, 271
		
\vspace*{0.5 truecm}
\begin{tabular}{|l|l|l|l|l|l|l|l|}\hline
\hline
\multicolumn{8}{|c|}{\bf Table 2: DECamERON program}
\\\hline
&&& Exp& time& kilo&secs&\\
\hline
Field&$\alpha,\delta$&&2012&2013&2014&2015&2016\\
&J2000&&Dec&Jan&May&Jul*&Jan$^\circ$\\
&     &&11 &12 &18 &23  &1\\\hline
New &&$g$&&&&3.0&\\
South &&$r$&&&&5.41&\\
-ern &22:32:56 &$i$&&& 5.28&13.2&\\
Field&--60:33   &$z$&&& 4.29&26.4&\\
(NSF)&&Y&&& 14.85& &\\\hline
Prime &5:55:07 &$g$&&&&& 2.0\\
Field&--61:21&$r$&&&&& 2.025\\
&&$i$&&&&& 1.65\\
&&$z$&7.2&7.8&&& 4.95\\
&&Y&8.4&7.8&&&11.55\\\hline
Polar &16:00:00&$r$&&& 2.5&&\\
Field&--75:00&$i$&&& 9.9&&\\
&&$z$&&& 18.48&7.5&\\
&&Y&&& 9.24&4.2&\\\hline
*also& $^\circ$also\\
28/12&  8/8\\
\end{tabular}
\vspace*{0.5 truecm}

\appendix
\leftline{\bf Appendix A}

Dark Energy Camera on the Blanco 4 meter telescope not only has the focal plane size the 1970s 4 meter telescopes were built for,
but also has good near infrared response. 
The goal of the DECamERON project\footnote{http://astronomy.swin.edu.au/$\sim$jmould/decameron.htm}
 is a deep wide field high redshift photometric survey to study large scale structure  and rare events. It reaches M* galaxies at redshift 6 at a wavelength of one micron, studying large scale structure and finding rare events.
The aim is
$not$ to compete with deeper narrow surveys like BORG (Trenti \etal 2010).
To maximize time allocation  flexibility, all 3 of these    
DECam deep fields are circumpolar.
Stacking of data is carried out by the DECam community pipeline (Valdes \etal 2014).
Here we report candidates for SNe at z = 4 in the Prime field. 
The light curves shown on the right are $z$ band.
Pipeline stacks are archived in 9 tiles of approximately 50$^\prime$ on a side.
These tiles form a 3 $\times$ 3 matrix on the sky with tiles 1,5,9 as the main diagonal.

The postage stamps are Y band images. Images of the same epoch are aligned in columns.
In Figures A1--3 we see supernovae that are bright in the first epoch, leaving only
the host galaxy in the most recent epoch.

\begin{figure*}
\begin{center}
\includegraphics[scale=.65, angle=-90]{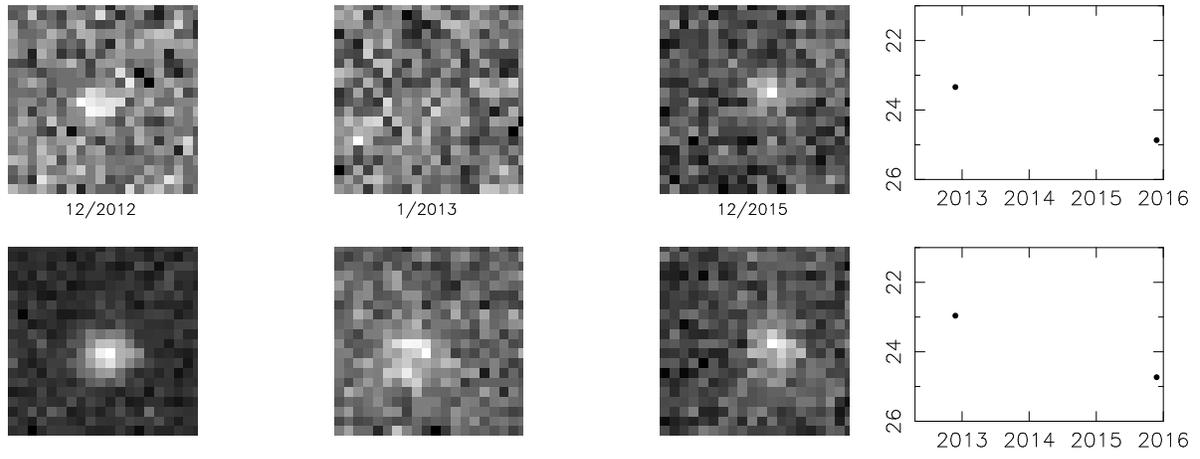}
\caption{Tile 2       z=4 candidates.}
\end{center}
\end{figure*}
\pagebreak
\begin{figure*}
\begin{center}
\includegraphics[scale=.65, angle=-90]{tile5.ps}
\caption{Tile 5       z=4 candidates}
\includegraphics[scale=.65, angle=-90]{tile8.ps}
\caption{Tile 8      z=4 candidates}
\end{center}
\end{figure*}

\end{document}